 \documentstyle[epsfig]{article}
\newcommand{\cl}[1]{\begin{center} {#1} \end{center}}

\begin{document}
\rightline{SBNCBS/NUP/96/02}
\vskip 0.3truecm
%\leftline{{\it Submitted to }}
\vskip 1.0truecm

\cl{\Large{Vlasov dynamics for fermions on phase space lattice}}

\vskip 1truecm
\cl{     S. Chattopadhyay\footnote{e-mail~: shila@tnp.saha.ernet.in} }
\vskip 1truecm

\cl{\it S.N. Bose  National Centre for Basic Sciences, Block-JD, Sector 3,
Salt Lake, Calcutta-700091, India}
\vskip 2truecm

\begin{abstract}

Within the framework of stochastic one-body approach a  simulation 
procedure to study the Vlasov dynamics for fermi system on phase 
space lattice is presented. To deal with fermions on lattice, 
the phase cell occupancy factors are taken to be either 1 or 0 in 
accordance with the Pauli exclusion principle. This has significant 
implications. First, the dynamical evolution does not alter initial 
temperature of the  system and secondly, at  finite temperature, 
the  proper  statistical  behaviour  related to  the fluctuations 
over the samples is ensured.  This method is applied for two distinct 
cases viz. the monopole  vibration of cold dilute nucleus and the 
evolution of ring shaped  matter  distribution  at finite temperature 
in the presence of radial flow. In the  latter case, fragment 
multiplicities  are found to be depend on the flow velocity.  To 
check the  reliability of the present  calculations simulations have 
been  performed with  different choices  of  the grid size. 

\end{abstract}


 \newpage

\centerline{{\bf 1. Introduction}}

In recent years, considerable efforts have been made to the study of nuclear 
dynamics within the framework of stochastic one body theories based on 
Boltzmann-Uheling- Uhlenbeck (BUU) equation. One of the major goals of such 
theories is to provide a consistent dynamical picture of fragmentation 
process near the fermi energy domain.  For the understanding of the 
fragmentation phenomenon at this energy regime, within the mean-field 
description, one plausible mechanism is to retain the effect of 
fluctuation in the two body collision term of the BUU equation. 
Several models and important methods of simulation based on Boltzmann- 
Langevin (BL) equation is proposed \cite{Ayi1}-\cite{Cha3} to incorporate 
stochasticity in to the deterministic transport equation. 

On the formal side, Ayik and Gregoire \cite{Ayi1} considered an extension of 
a one body transport model by incorporating higher order correlations in 
the equation of motion in a stochastic approximation. Later, in a development 
of the numerical procedure, the evolution of moments of the phase space 
density distribution is carried out by calculating the correlation matrices 
at every time step. This in turn, allows to construct a trajectory or the 
new phase space density in the ensemble space \cite{Ayi2}. In principle, by  
this manner one can specify the new density uniquely provided an infinite 
number of multipole moments are taken in to account. But, in practice, the 
evolution of phase space density is considered in the terms of limited 
number of multipole moments so that simulation  of the reaction 
dynamics in 3 dimension physical space can be performed (for further 
details see the review \cite{Ayi3}). However, in another approach, Randrup 
and Remaud \cite{Ran1} recast essentially the same problem within the 
framework of Fokker-Planck equation. In this method the stochasticity is 
directly incorporated into the basic two body scattering process. In the 
numerical implementation  of this method, the phase space density 
$f({\bf r},{\bf p})$ around a point (${\bf r},{\bf p}$)is represented 
on the phase space lattice having elementary volume 
$\Delta {\bf s}$(=$\Delta{\bf r}\Delta{\bf p}/h^D$, D being the 
dimension of the physical space). The phase cell occupancy factor 
$n({\bf r}, {\bf p})$ is given as 
$f({\bf r},{\bf p})=n({\bf r},{\bf p})\Delta{\bf s}$. Due to the individual 
collisions, net flow of phase space density from one of the elementary cells 
to the others is considered to be a fluctuating quantity characterised by a 
Gaussian noise \cite{Ran2}. As a result, in this semi-classical treatment, 
the change of the phase cell occupancy factor $n({\bf r},{\bf p})$ can occur 
in a continuous manner so that $n({\bf r}, {\bf p})$ may attain a unphysical 
value either greater than 1 or less than 0. This feature, in one hand, 
allows us to incorporate efficient algorithm for finding the solution of 
Vlasov part of the stochastic BUU equation in grid space as described in 
details in Ref. \cite{Ran4}. 
On the other hand, one needs a provision for smearing of density forcefully 
within the neighboring phase cells. Although the 
application of this lattice model in 3D realistic situation is still a 
formidable task, it has been however, is rigorously tested for simpler cases  
\cite{Ran2} and also applied to study the growth of density fluctuation due 
to the thermal agitation of uniform nuclear matter \cite{Ran3},\cite{Ran4}. 
Simpler 
algorithm based on `test particle approach' is also put forward \cite{Ran5} 
which becomes an important tool of studying different aspects of fluctuation 
phenomena \cite{Tor1}. In case of heavy-ion collision, this model offer an 
explanation of the observed large fluctuation in the projectile and target 
like masses near peripheral collisions \cite{Tor2}. These apart, within the 
framework of Vlasov equation some interesting investigation associated with 
the growth of density fluctuation subjected to a given initial condition has 
also been reported \cite{Bal}.

Let us concentrate on a particular approach for inclusion of fluctuation 
within the BUU dynamics as developed in Ref. \cite{Cha1} and  \cite{Cha2}. 
Accordingly, the one body phase space of the system is 
discretise into separate small cells of size $\Delta {\bf s}$  
and the ensemble space is generated by  ${\cal N}$ such samples of phase space.  
To describe fermions on the phase space lattice, each of the enumerated 
phase cell (say $i$) is filled up in such a way that every phase cell $(ik)$ 
($k$ run from 1 to ${\cal N}$) is either completely filled or vacant. For a 
given value of ensemble average occupancy factor $\langle n\rangle_i$ for each 
$i$ we can fill ${\cal N}$ number of boxes by identical particles in different 
ways  provided the value of $\langle n\rangle_i$ is not either 1 or 0. In a 
broader sense, the underlying process responsible for this typical filling 
up procedure is known as dichotomic stochastic process. The sample mean and 
variance of the said fluctuation is given by 
\begin{equation}
\langle n\rangle_i={1\over{\cal N}}\sum_{k=1}^{\cal N} n_{ik}
~~~~~~~~~~{\rm and}~~~~~~~~~~~\sigma_i^2=\langle n\rangle_i(1-
\langle n\rangle_i)
\end{equation}
It is to be noted that relation (1) is {\it independent} of the size of 
$\Delta{\bf s}$. Another important feature is that, whatever be 
the simulation method used to study  the evolution of phase space density 
or the occupancy factor $n_i(t)$  eq. (1) always holds. Even if 
we consider the evolution of hot ($T\neq 0$) fermi gas through Vlasov equation 
no matter whether spatial uniformity retains or the clusterisation occurs,  
this statistical criteria is always satisfied separately for every phase 
cell $i$. To study the effect of the two body collisions on the evolution of 
one body phase space density, in semi-classical approximation one usually 
employs the BUU type collision term to the evolution equation. To account for 
the fluctuations associated to the individual collisions among the particles 
a method of simulation is proposed in Ref. \cite{Cha1}. in a consistent 
manner so that the proper balance between the outflow and the inflow term in 
the collision part of BUU equation is maintained on the sample average 
description. As a result, the presence of of the collisions change 
$\langle n\rangle_i(t)$ of the state in a different way (unlike to that 
provided by the Vlasov equation) so that, at the end 
$\langle n\rangle_i(t\to\infty)$ represents a equilibrium situation with 
certain local (or global) temperature and density (see Ref. \cite{Cha1}). 
Within this formalism the concept of single or mean trajectory can not be 
retained. For $\Delta{\bf s}\to 0$
the exact solution of BUU (or the Vlasov) equation is recovered for the 
average description. However, for the fluctuation, relation (1)  always holds 
which is independent of the choice of $\Delta {\bf s}$. This is the key 
result. As long as we stick to the above mentioned description, namely, 
the values of phase space occupancy factors are allowed to changes in discrete 
manner (either from 0 to 1 or from 1 to 0 otherwise) 
the proper measure of fluctuations over the samples are been maintained 
throughout. As a consequence, the main effort of the simulation procedure is 
directed at predicting the proper evolution of the ensemble average 
occupancy at each phase cell only and to check the consistency, 
we may compare the simulation results with different choices of 
$\Delta {\bf s}$. Need less to say that as we decrease the size of 
$\Delta x$ or that of ($\Delta p$) the storage requirement increases 
substantially. Moreover, unlike the test particle approach, in our case, 
we have to consider not only the particle states but also the hole states. 
As a result the memory requirements increase tremendously 
with the dimension of the physical space. It is important to note that  
we treat our occupancy factors $n_i$ as binary numbers which essentially 
reduces  the storage requirements to a certain extent (see \cite{Cha1}). 
However, to perform a 3D calculation for a realistic situation one have 
to choose bigger size of both $\Delta x$ and $\Delta p$. 
Therefore, to find out an optimal values of grid sizes, our strategy 
is to compare the results with different choices 
of $\Delta {\bf s}$ in 2D.  In this  situation we have a greater freedom of 
such choices.

As has already been stressed that, our  main concern here is to generate proper 
evolution of sample average 
distribution function with reasonable accuracy. For this purpose, therefore, 
investigation can be persuaded within the framework of Vlasov equation. 
In addition, it provides an opportunity to explore  
how reliably the time evolution can be described within our formalism where 
the occupancy factors can take only two discrete values.  As a test case 
we have already studied 
the problem of spinodal decomposition of nuclear matter \cite{Cha2}. 
However, for finite nuclear system one has to check the reliability of this 
new method. 

In order to consider propagation of 
phase space density distribution we follow Nordheim's approach, which is 
closely related to the well-known test 
particle method of solving the Vlasov equation \cite{Cha2}. A filled unit 
phase cell here is considered to be a particle. The time evolution of these 
particles can be obtained  by the standard leap-frog routine through the 
following equation 
 \begin{eqnarray}
  x^i_k(t+\Delta t)&=&x^i_k(t)+{\cal C} p^i_k(t),\nonumber\\
  p^i_k(t+\Delta t)&=&p^i_k(t)+
{{\Delta t}\over{\Delta p}}\nabla\big(\tilde{\cal U}[\rho(r,t)]\big)_k,\\
~~{\rm where}~~~~~~~~~~~~~ {\cal C}&=&{{\Delta t}\over m}
{{\Delta p}\over{\Delta x}}.\nonumber
\end{eqnarray}
 Here $({\bf x}^i,{\bf p}^i)$ represents the co-ordinates of the $i$th 
particle in the phase space grid which can take only integer values. 
It may be noticed that to maintain 
uniformity of the solution over the position space in the case of free 
streaming gas one has to choose the value of ${\cal C}$ to be unity. 
Therefore for a given  grid size the value of time step $\Delta t$ 
remains fixed. This ensures that two particles do not appear in the same 
spatial cell due to the integer truncation during the evolution. 
This feature then accounts the proper incorporation of Pauli principle 
in case of Vlasov propagation of $n_i$. The importance of this issue has 
been recently explored \cite{Rei} by the uses of 
'test- particle' procedure. Using a well-controlled Vlasov algorithm, it is 
revealed that the error which arises primarily due to the improper treatment 
of Pauli principle, drive the collection of test-particles to a state of 
classical equilibrium. The manner in which fermions 
are treated within our formalism such a situation never arise  which can be 
understood as follows. For a temperature $T=0$ , the system is considered to 
be a statistically pure state so that all samples are identical. Due to 
the deterministic nature of the Vlasov 
equation they are remain at the same state. Because the fluctuation 
$\sigma^2_i$ is always equal to zero for every cell $i$, the system does not 
acquire any temperature. The error which arises in the simulation  does not 
however, excite the system to a different temperature. For the 
calculation of gradient term of the effective potential $\tilde{\cal U}_i$ in 
equation (3) we introduce a Gaussian folding function having width $\sigma_r$  
so that 
\begin{equation}
\tilde{\cal U}_i={1\over{{\cal N}_g}}\sum_j 
{\rm e}^{-{{\big({\bf r}_i-{\bf r}_j\big)^2}\over{2\sigma^2_r}}}
{\cal U}({\bf r_j})
\end{equation}
where ${\cal N}_g=\sum_j{exp}\big[-({\bf r}_i-{\bf r}_j)^2/(2\sigma^2_r)]$, is 
the normalization constant and is independent of $i$. The bare potential 
${\cal U}[\rho({\bf r}_j)]$ on the grid may be calculated by using 
standard Skyrme interaction as used in earlier calculations \cite{Cha2}. 
To reduce the truncation error that arises in finding the time 
development of momenta of the particles on grid a method is invoked in 
Ref. \cite{Cha2}. To provide  better energy 
conversion one may incorporate this correction in the above algorithm. 

In the present paper we apply this method of solving nuclear Vlasov equation 
in two specific situations. In section 2 we study a typical case of monopole 
vibration of cold isolated nucleus and in section 3 we study the evolution 
of hot circular ring in presence of radial flow. In this case, inhomogeneity 
that resides in the initial state of the system, grows with time which 
results in the formation of fragments; both of these calculations are 
performed  in 2D physical space. Finally in section 4 we summarize our 
simulation result. 

\centerline{{\bf 2. Monopole vibration}}

In this section we shall present simulation results of  monopole 
vibration of a cold dilute nucleus. To check the reliability of our 
calculation we repeats our calculation for different choices of grid 
width $\Delta p$ and $\Delta x$. In this situation we consider the 
evolution of a single sample at temperature $T=0$. The phase space 
distribution of the particles is taken to be spherical in both 
position and momentum space with radii $R_m$ and $P_m$ so that density 
profile $\rho(R)=\Theta(R-R_m)$, and the magnitude of $P_m$ is scaled 
to $\eta P_F$. It is to be noted that at $R=R_m$ the density distribution 
undergoes a discontinuous jump. However, due to the introduction of the 
folding function in the expression of effective potential as shown in 
eq.(3), considerable  smoothening in the single particle potential can 
be ensured. Therefore, the effect of the smooth tail in the matter 
distribution is automatically taken care of. It may be also  mentioned 
that self-consistency between mean field, within which the particles 
moves and the density at the initial state cannot be achieved in this 
manner. In the context of semi-classical approximation, the present 
problem of monopole vibration can be treated in a consistent way without 
imposing the scaling approximation of momentum sphere \cite{Tor3}. 
Within the test particle approach, when the scale parameter $\eta$ is 
taken to be either very low or very high compared to unity large 
oscillations in density can be predicted. On the other hand in case of 
small oscillation where the value of $\eta$ is chosen to be close to 1, 
it is observed that oscillation is dies down very rapidly, as a result 
the energy in the collective mode is transferred into the random motion 
\cite{Rei},\cite{Ber0}. Therefore, it seems appropriate to direct our 
study for the case of small oscillations only. For a proper description of 
initial state at temperature $T=0$ on grid of finite size, $\eta$ can not 
vary in a continuous manner. In particular, for $\Delta p= 58 $MeV/c and 
74 MeV/c we have chosen the values of $\eta$ to be $\simeq 0.88$ and 
$\simeq 0.75$ respectively. For a given value of fermi momentum 
$P_F=260$MeV/c, these values  of $\eta$ cannot be made any further 
closer to 1. Here, we present the simulation for two chosen values of 
$\Delta x =$ ${1\over 3}$ fm and 0.5 fm. The time steps $\Delta t$ as 
decided through the relation ${\cal C}=1$ mentioned above are different 
for different choices of $\Delta p$ and $\Delta x$. Accordingly, the 
values are ${\Delta t}\simeq$ 
5.4 (for $\Delta p=58 {\rm MeV/c}$, and $\Delta x={1\over 3}$ fm), 
7.1 (for $\Delta p=58 {\rm MeV/c}$, and $\Delta x={1\over 2}$ fm), 
4.1 (for $\Delta p=74 {\rm MeV/c}$, and $\Delta x={1\over 3}$ fm), 
6.5 (for $\Delta p=74 {\rm MeV/c}$, and $\Delta x={1\over 2}$ fm). 

Firstly, we observer that Pauli exclusion principle is obeyed within 
our simulation procedure or, in other words, no two particles access 
the same phase cell in their process of evolution for all the studied. 
This feature may be related to the observation that the symmetry in 
both the spatial and the momentum distribution of the initial state is 
preserved accurately throughout the evolution. From the point of view 
of symmetry argument, this is a stringent case where the full spherical 
symmetry in 2D should be considered. Hence, all the multipole moments  
of the said distributions (both for momentum and for position) remain at 
zero. To get an overall idea about the evolutionary pattern we plot in 
Fig. 1(a) and (b) the time development of mean square radius of the 
matter distribution for different values of scale parameter $\eta$ and 
for comparison, we also plot the same for different choices of 
$\Delta x$. It is to be noted that for different choices of $\Delta p$ 
and $\Delta x$ we cannot prepare exactly identical states of the 
nucleus. Consequently, there arises a slight variation in the 
binding energy and also in the rms radius of the initial state.
With $R_m=6$ fm mass number of the nucleus is found to  lie within 
the range 64$\pm$2. Simulation of the Vlasov equation being an initial 
value problem, the amplitude and the initial phase of the the 
oscillation in the rms radius (or the density) depend  on the shape 
of the  potential profile $\tilde{\cal U}(r)$ of the the initial state. 
However, the time period of the oscillations for different values of 
spatial and momentum grid size is seen to be almost the same. As we 
mentioned above, the error that arises mostly due to the incorporation 
of the leap-frog routine for finding the time evolution of phase space 
density does not initiate dissipation. On the other hand, this error 
may attribute a certain amount of mismatch between the momentum 
distribution and the profile of the effective potential.  As a result 
the particles those start from  one side of the nucleus may leak off 
from the boundary on the other side of it. The lack of the 
self-consistency in the initial state partly responsible for such 
'evaporation' which is not expected at T=0. beyond the time t=70 fm/c, 
which is of the order of the transit time of the particles, evaporation 
starts. This feature affects the measurements of $r^2_{rms}$. The time 
average values of this quantity show an upward trend. Hence, to reduce 
this effect, we set up a criteria namely, the particle which lie beyond 
a sphere of 10 fm radius are considered to be evaporated, and are not 
employed in the calculation of $r^2_{rms}$. We observe that within time 
$\simeq$250 fm/c, the mass number of the nucleus is reduced by 3-4.5. 
However, the evaporation rate reduces considerably as we decrease the 
size of the spatial grid or the size of the time step.

It is an well-known fact that the energy conservation is not realized 
very accurately within the particle method for finding the solution 
of partial differential equation. Apart from some simpler situations, 
mainly due to the storage problem, it is very difficult to provide a 
solution of Vlasov equation on phase space lattice incorporating 
superior methods such as described in Ref. \cite{Ran5}. Although we use phase 
space grids for description of fermions, but in order to invoke the 
essential properties of the underlying nature of the fluctuation as 
given by equation (1), phase space occupancy cannot be changed in a 
continuous manner. As a result, in spite of our use of large number 
of particles for description of normal nuclear matter in a unit box,  
the magnitude of total energy per particle $E/A$ at every time step 
cannot be  preserved very accurately. The error arises mainly due to 
the truncation of momenta $p_i$. An upper limit of the estimated 
error in the measurement of total kinetic energy per particle may be 
given by $(\Delta p)^2/2m$ which is $\simeq 1.5$ MeV for 
$\Delta p=60$MeV/c. In Fig. 1(c) and (d) we plot the evolution of E/A 
for different situations. Apart from some fluctuation, within 
the time of 200 fm/c the error of E/A lies within the 1 MeV. Beyond 
it the slop of E/A curve rises up further. It is clearly seen that 
overall slope of this curve decreases as we decrease the time step 
of the simulation.

\centerline{{\bf 2.Fragmentation within Vlasov dynamics}}

In this section we study the evolution of nuclear matter  
having a shape of a circular ring in the presence of radial flow at a 
given finite temperature T $\not=0$. This particular situation seems 
to be very appropriate for investigation of the dynamical aspects 
of fragmentation at central collision. Several theoretical calculations 
revealed that in later stage of nearly central collision, beyond an 
energy $\simeq$ 40 MeV/A, the post collision nuclear complex takes a 
shape of a torus \cite{Ber1} and the fragments may appear by the 
breaking up of such a structure. Using statistical models, investigation 
has also been performed to provide an quantitative estimate about the size 
and multiplicities of the fragments and their correlation with kinetic 
energies \cite{Pal}. From  the more general framework of 
Boltzmann- Langevin model, such studies have also been carried out in the 
case of 2D physical space \cite{Cha3}. This shows that due to the presence 
of strong flow along the transverse direction an elongated ring like 
structure is formed with larger deposition of matter along the transverse 
direction and the size or mass of the fragments depends on the speed of 
stretching of such structure. Thus, the size of the fragments depends on the 
magnitude of the radial flow. To get a rough idea about the expected outcome 
from a 3D calculation one may rotate such a two dimensional object around its 
symmetry axis along the direction of collision.  This leads essentially 
to a bubble like structure. However, the mass around its boundary is 
definitely not uniform. Along the transverse plane a large deposition 
of matter having a toroidal shape is expected. Due to the presence of 
relatively large radial flow in this plane the torus may break quickly 
into several fragments. To mimic this situation as predicted from our 
earlier calculation \cite{Cha3}, here we consider the evolution of ring like 
structure in 2D in the presence of radial flow.

In the present cases, we take the value of average density 
$\langle\rho\rangle$ to be ${1\over 2}\rho_0$ which is considered to be 
uniform within the annular region between two concentric circles of radii 
5 fm and 15 fm. For proper statistical description of the initial state 
at temperature T$\not=0$ we prepare the samples as described in 
Ref. \cite{Cha2}. 
The  average occupancy factor $\langle n\rangle_i$ over the sample is 
taken to be a fermi distribution with an  appropriate value of chemical 
potential $\mu(\langle\rho\rangle ,T)$ at a given temperature T. In 
order to  determine the momentum distribution at every spatial cell
we prepare a set of 100 samples so that sample variance of the total 
number of particles at each spatial cell is given by the relation 
$\sigma^2(N)=\sum_i\langle n\rangle_i(1-\langle n\rangle_i$ where 
the sum is on the momentum state only. Next, this samples are distributed 
randomly at every spatial cell in such a manner so that the sample average 
values of the density at every spatial cell within the annular region 
remains at ${1\over 2}\rho_0$  and average momentum distribution  
is represented by identical fermi distributions as mentioned above. 
By this manner we can introduce the spatial fluctuations in density of 
about 10\% within each of the samples at temperature $T=7$MeV. To 
incorporate radial flow in the initial state one has to provide a boost 
to the  the momentum distribution  of every spatial cell $i$ so that 
the magnitude of the boost momentum is given by $mv_{fl}$ and its 
direction at every spatial cell is decided through the relation 
$\tan(\theta_{fl})=y_i/x_i$. 

Let us concentrate on the simulation result. Due to the presence of the 
radial flow in the initial state, the size of the central hole increases 
gradually with time. In addition, the spatial fluctuation that resides 
in the initial state grow with time which leads to the appearance of 
fragments in the later stages of the evolution. The energy associated 
with the radial flow is continuously transferred into the internal energy 
so that the rate of recession decreases gradually. Clusterisation is 
observed even in the average level of the spatial density distribution. 
This is a general feature of the evolution that we observe here. To check 
the reliability of our calculation we have performed our simulation for 
different values of $\Delta {\bf s}$. For the first case, we chose the 
values of $\Delta p=58$ MeV/c and $\Delta x={1\over 3}$ fm so that the 
number of particles needed to describe normal nuclear matter within a box 
of size 1 fm is given by 603 whereas this value is 148 for the choice of 
$\Delta p =74$ MeV/c and $\Delta x={1\over 2}$ fm. In Fig. 2 and 3, we 
plot the sample average density on the spatial grid at two regime of the 
dynamics for two given initial values of flow velocity ($v_{fl}=$ 0.12c 
and 0.9c) with the smaller size of $\Delta {\bf s}$. Initial transients 
in the density fluctuations (Fig. 2(a) and 3(a)) ultimately leads a 
steady pattern. The clustered nature of this pattern (Fig. 2(b) and 3(b)) 
yields evidence of the trend towards fragmentation. Later on, the relative 
distance between the clusters gradually increases. It is also to be noted  
that similar such patterns is also observed in our simulation for other 
choices of $\Delta {\bf s}$. The  crucial role of the initial flow on this 
pattern is evident from these figures. Fragmentation starts after this 
steady pattern appears in the matter distribution. Therefore, one may 
safely guess at the average number of the fragments produced. To identify 
fragments in every individual samples we use the `spanning tree method' 
where  We take the boundary of a fragment on the spatial grid to extend 
to the point where matter density reaches a value lower than 0.11$\rho_o$. 

The mode of fragmentation can be split into two distinct regime. Firstly, 
due to the continuous  stretching fluctuations grows to a certain extent, so 
that the ring shaped structure initially breaks up with at least one large 
chunk which very quickly undergoes fragmentation. The average size or the 
mass of such clusters is determined by the initial conditions, primarily the 
magnitude of the flow velocity $v_{fl}$. Later, due to evaporation the mass 
of the fragments reduces steadily. Although the time scale of the 
fragmentation process depends crucially on $v_{fl}$ and also on temperature, 
these distinguishing features of the dynamics are present in every case, 
this  can easily be recognised from the study of time evolution of fragment 
mass abundances. To quantify, in Fig. 4 we plot the frequency distribution 
versus fragment mass at different times for 100  samples. To make a direct 
comparison, the time evolution of fragment mass abundances for two distinct 
cases are shown in the same panel. Depending upon the initial values of 
$v_{fl}$ this distribution peaks around two different values of mass number.  
Even at early stages of the fragmentation, we observe this characteristic 
feature of the dynamics. However, because of evaporation both the peaks 
shift towards smaller values of mass number in course of time.  
It is to be mentioned that we could not provide a proper description of the 
production of light particles such as proton, neutron, deutron, triton etc. 
In the current scheme even more than  30\% of the initial mass simply 
evaporated out at the earlier stages of the evolution (when production of 
such particles is expected) without producing any fragments. To get an 
idea about how far we can relay on our calculation for the details of the 
fragmentation dynamics, in Fig. 5 
and 6 we compare our  results on time evolution of mass abundances that are 
shown in Fig. 4 with the same date which are extracted from the simulation 
with other choices of $\Delta{\rm s}$. As we increase the grid size, the 
rate of evaporation increases. Apart from this, the nature of the spectra 
does not change which can also be seen from the plot of cumulative 
frequency versus fragment mass given in Fig. 6 for higher values of $v_{fl}$. 
To get an idea about the promptness of the fragmentation process, we plot the 
time evolution of average number of fragments per sample in Fig. 7. The time 
beyond which fragments appear, is defined as the onset time $t_{onset}$. 
Beyond this time, the number of fragments per sample ultimately saturates to a 
certain steady value. The time that is needed to reach this saturation (in 
other words the fragmentation time $t_{frag}$) can be estimated from this 
curve. For $v_{fl}=0.12c$ the estimated value of $t_{frag}\simeq 50$fm/c which 
is quite small in comparison to the value of $t_{onset}$($\simeq 125$ fm/c).  
For lower values of $v_{fl}$, one has to wait for a much longer time 
$\simeq$ 200 fm/c to observe fragments. In this case, it is  
observed that the estimates of $t_{onset}$ as well as the total number of 
fragments changes only to be slightly for different choices of $\Delta{\bf s}$. 

{\centerline{\bf 4. Summary}}

In order to treat fermions in the same way as provided by the Boltzmann -
Langevin model, we introduce in this paper a simulation method of 
Vlasov dynamics on phase space lattice. This simulation procedure is 
designed in such a manner that the Pauli principle can be ensured 
very accurately and proper statistical criteria related to the fluctuation 
of phase space density (as given by eq.(1)) can be fulfilled. This becomes 
an important issue, specially, for the study of growth of fluctuations  
in hot fermi systems. In the context of Vlasov dynamics, this feature guarantee 
an evolution of isothermal samples.  For illustration of this method we 
consider two specific cases namely; monopole vibrations of a cold dilute 
nucleus and the evolution of ring shaped matter distribution in the presence 
of thermal noise. 

For the first case, it is observed that within our simulation 
scheme,  the symmetry that resides in the initial 
state of the spherical nucleus is preserved under Vlasov propagation of 
phase space density. This  result is verified further from the simulation with 
different choices of phase cell size ${\Delta{\bf s}}$. However, in spite of our 
use of a large number of particles for simulation, the magnitude of 
total energy per particle of the system can not be retained with enough accuracy.  
The problem mainly arises due to the application for large time step of the 
leap-frog routine. In order to provide an energy conserving evolution 
we can not take the most obvious option (as adopted in the test particle 
approach of simulation) of choosing the time step to be arbitrarily small. 
In such cases we invariably face the problem of incorporating pauli principle 
in  the dynamics. 

It is to be noted that at zero temperature, the system is defined uniquely 
by a single trajectory (or sample). On the other hand, in case of finite  
temperature,  one needs a collection of samples for proper description of 
the system to a certain accuracy one needs a collection of samples , the 
degree of which should increase with the size of the  collection.  
In this situation, the concept of a single or mean trajectory does not 
make any sense. The initial fluctuations in density within the samples 
arises solely due to thermal agitation. If we consider a situation of thermal 
equilibrium, the amplitude of the density fluctuations is 
decided by the magnitude of temperature and sample average density. 
Within our method of simulation, it becomes possible to provide a proper 
description of noise consistent with the average  properties of the system 
even at time $t=0$. Within the Boltzmann-Langevin model, because of the 
presence of the fluctuation collision term (or the source term) the details 
of the evolution do not depend very much on how one prepare the initial 
samples. The fluctuations  associated with the initial state may appear 
in a additive manner to the dynamics which presumably, cannot be true 
in the case of Vlasov dynamics. Due to the deterministic 
nature of the evolution, the inhomogeneity that exists initially within every  
sample, develops independently with time. To asses 
its role in the dynamics, an investigation was carried on by the authors of 
Ref \cite{Bal}. Using a very accurate algorithm for the simulation of Vlasov 
equation they found that the neighbouring particle trajectories diverge 
exponentially with time, concluding, that the  associated dynamics may 
be characterised by classical noise. As the situation demands, we perform
our simulation with a different initial temperature of 3.5MeV. Although 
the essential features remains to be unaltered as seen by the observation 
of the similar evolutionary pattern (as given by Fig. 2 and 3.), however, 
the fragments appear very late in time. For $v_{fl}=0.12c$, the onset time 
of the fragments takes a value $\simeq200$fm/c. We also repeat our 
calculations with different sets of initial samples. 
The observed differences may arise due to the fact that the sample sets  
that are prepared with different choices of $\Delta{\bf s}$  are not identical. 

In a realistic situation of ion-ion collision, the formation of structure of 
toroid shape has been predicted in different calculations. 
The usual statistical models that are employed to study the fragmentation 
of a hot residue has no provision to consider the dynamical features of 
collective radial flow. 
Within a less complicated scenario of two dimension, we investigate 
the role of radial collective flow on the dynamics of fragmentation of 
circular ring, a 2D counterpart of the torus. 
We also emphasis on the fact that the results of our simulation 
does not depend on the choices of grid size. This is an important 
aspect, as otherwise, one can not provide a proper estimate of fragment 
multiplicities from a dynamical calculation.

\newpage
\begin{figure}
\begin{center}
\epsfig{file=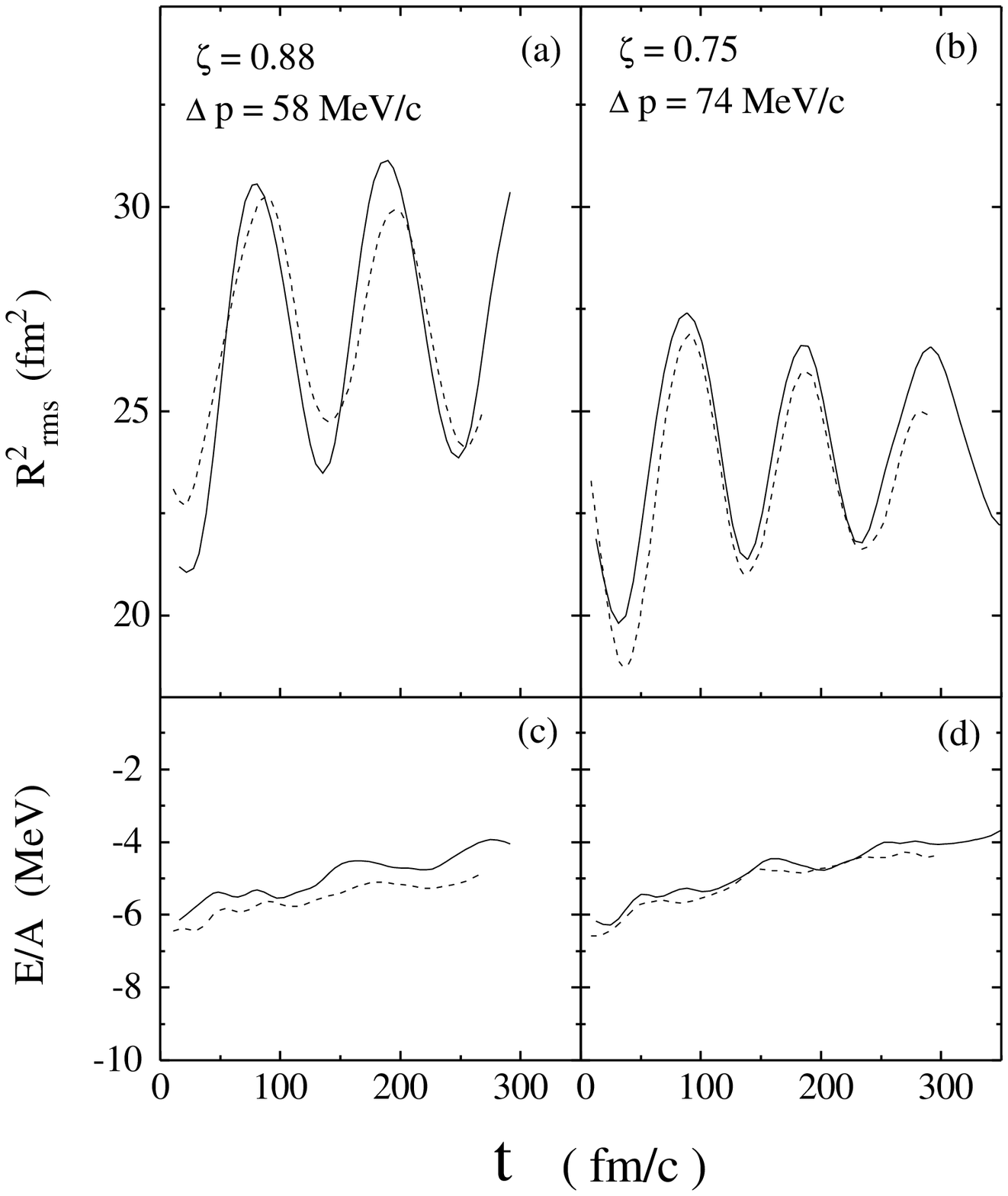,width=0.8\textwidth}
 \vspace{-1.0cm}
\end{center}

 {\bf Fig. 1} \\
Time evolution of mean square radius $r^2_{rms}$ and energy per particle 
E/A of monopole vibration of an isolated nucleus  is shown for different 
values of momentum scale factor $\eta$. The solid and dotted curves represent 
the simulation results with $\Delta x=$ ${1\over 3}$ fm and 
${1\over 2}$ fm respectively.
\end{figure}

\newpage
\begin{figure}
\begin{center}
\epsfig{file=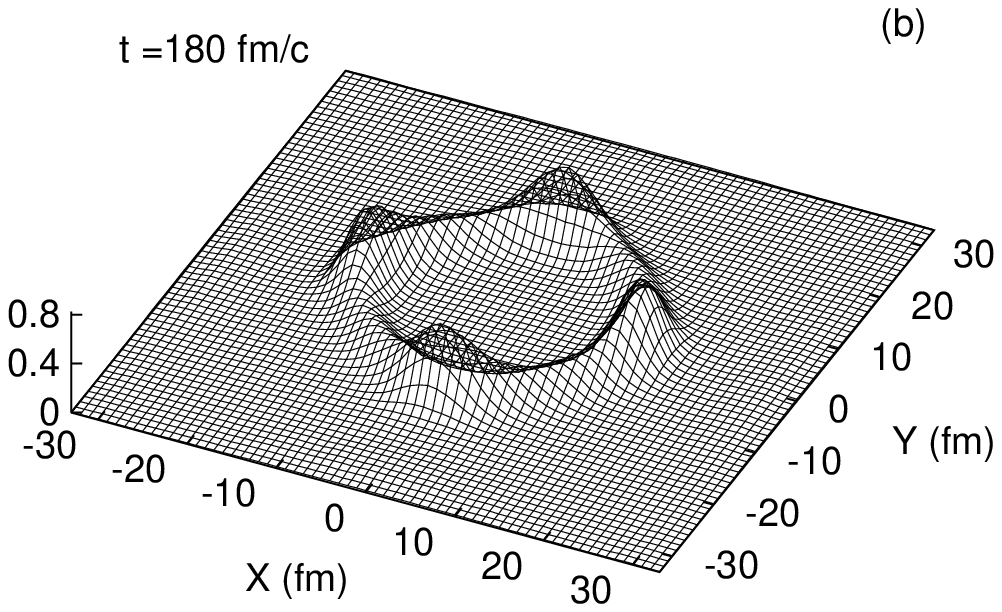,width=0.8\textwidth}
 \vspace{-1.5cm}
\epsfig{file=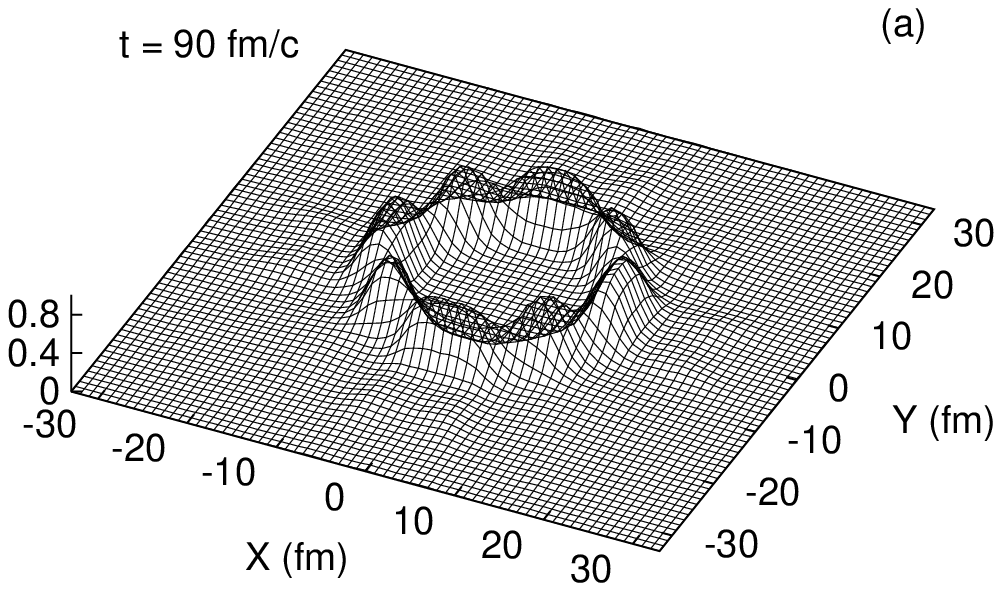,width=0.8\textwidth}
\end{center}

{\bf Fig.2} \\
 Spatial variation of the sample averaged density distribution 
 $\langle\rho(x,y;t)\rangle$ in unit of $\rho_0$ is exhibited for the 
 initial flow velocity $v_{fl}$= 0.09c. The simulation results are with 
 $\Delta p=$ 58 MeV/c and $\Delta x={1\over 3}$ fm. 
\end{figure}

\newpage
\begin{figure}
\begin{center}
\epsfig{file=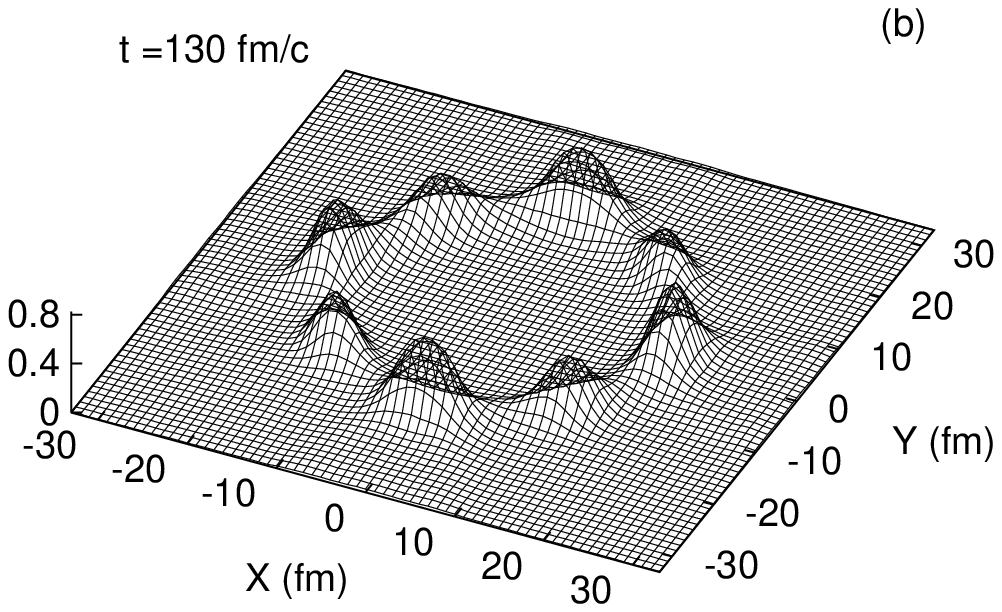,width=0.8\textwidth}
 \vspace{-1.3cm}
\epsfig{file=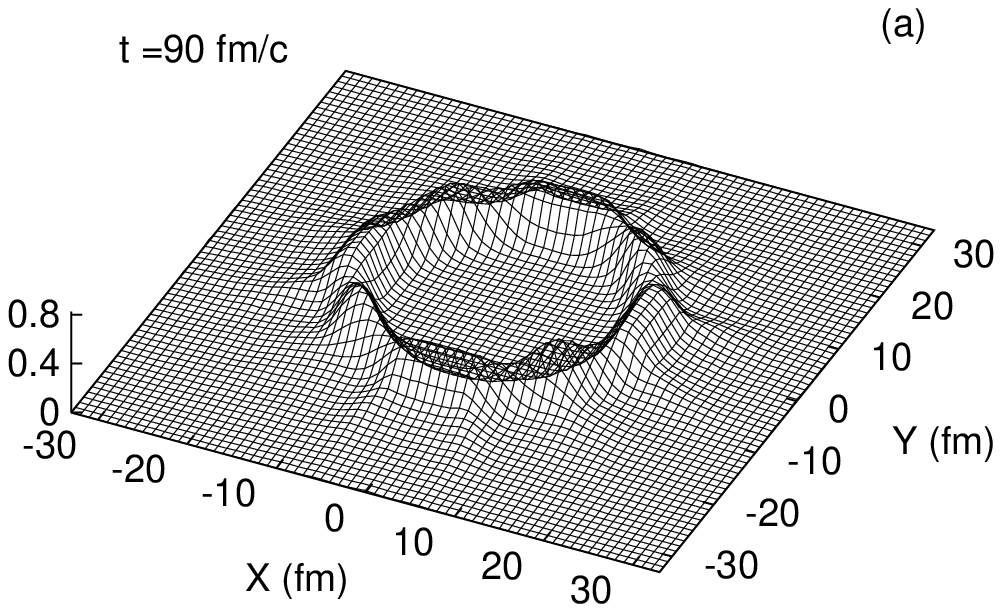,width=0.8\textwidth}
\end{center}

{\bf Fig.3} \\
Same as that of Fig. 2 with $v_{fl}=$ 0.12c. 
\end{figure}

\newpage
\begin{figure}
\begin{center}
\epsfig{file=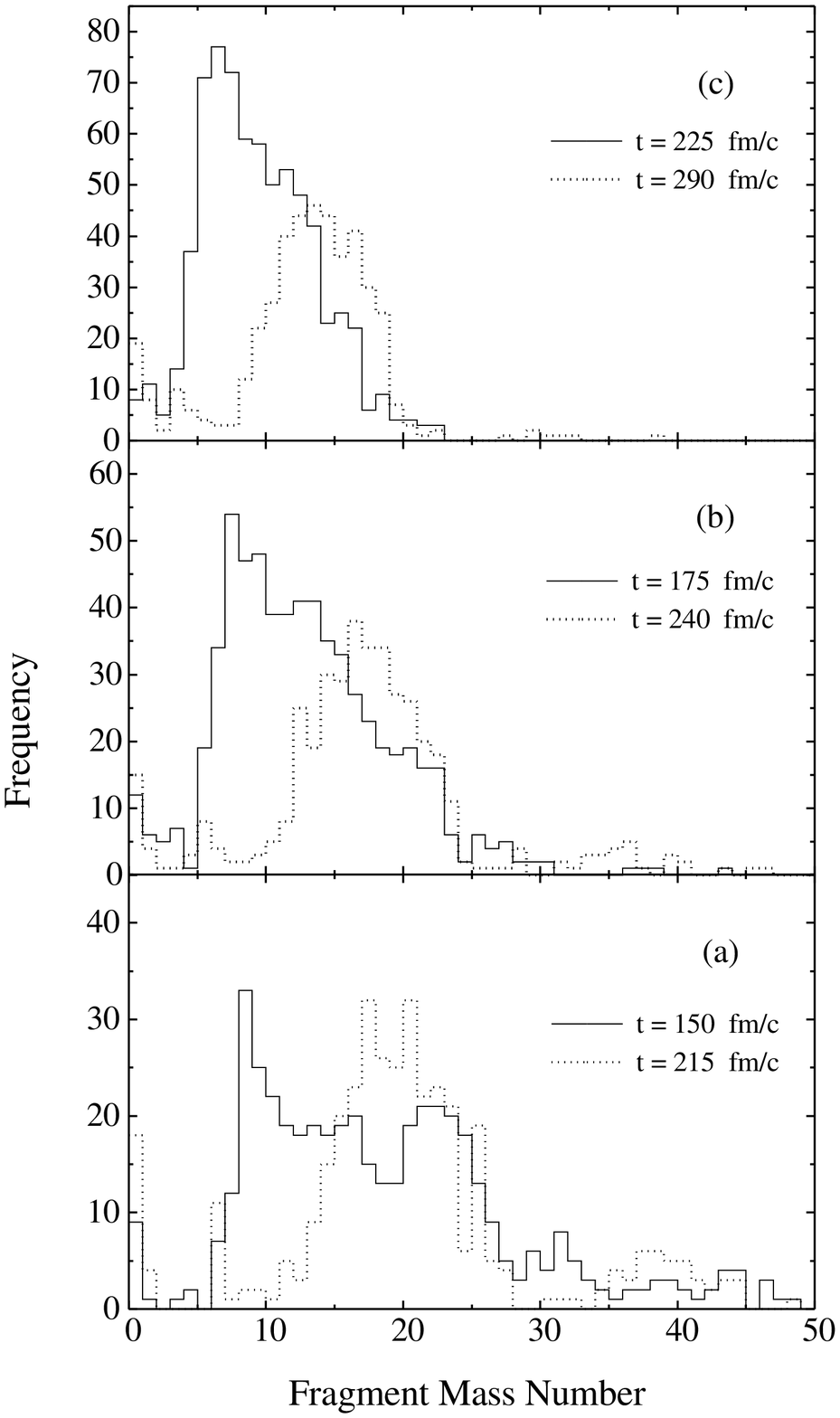,width=0.8\textwidth}
 \vspace{-1.0cm}
\end{center}

{\bf Fig.4} \\
Time evolution of fragment mass abundances are plotted for two 
different values of initial flow velocity $v_{fl}=$ 0.12c (solid histogram) 
and 0.09c (dotted histogram).  Simulation results with 
$\Delta p=$ 74 MeV/c and $\Delta x= {1\over 2}$ fm are shown for different 
times in panels (a)-(c). 
\end{figure}
\newpage
\begin{figure}
\begin{center}
\epsfig{file=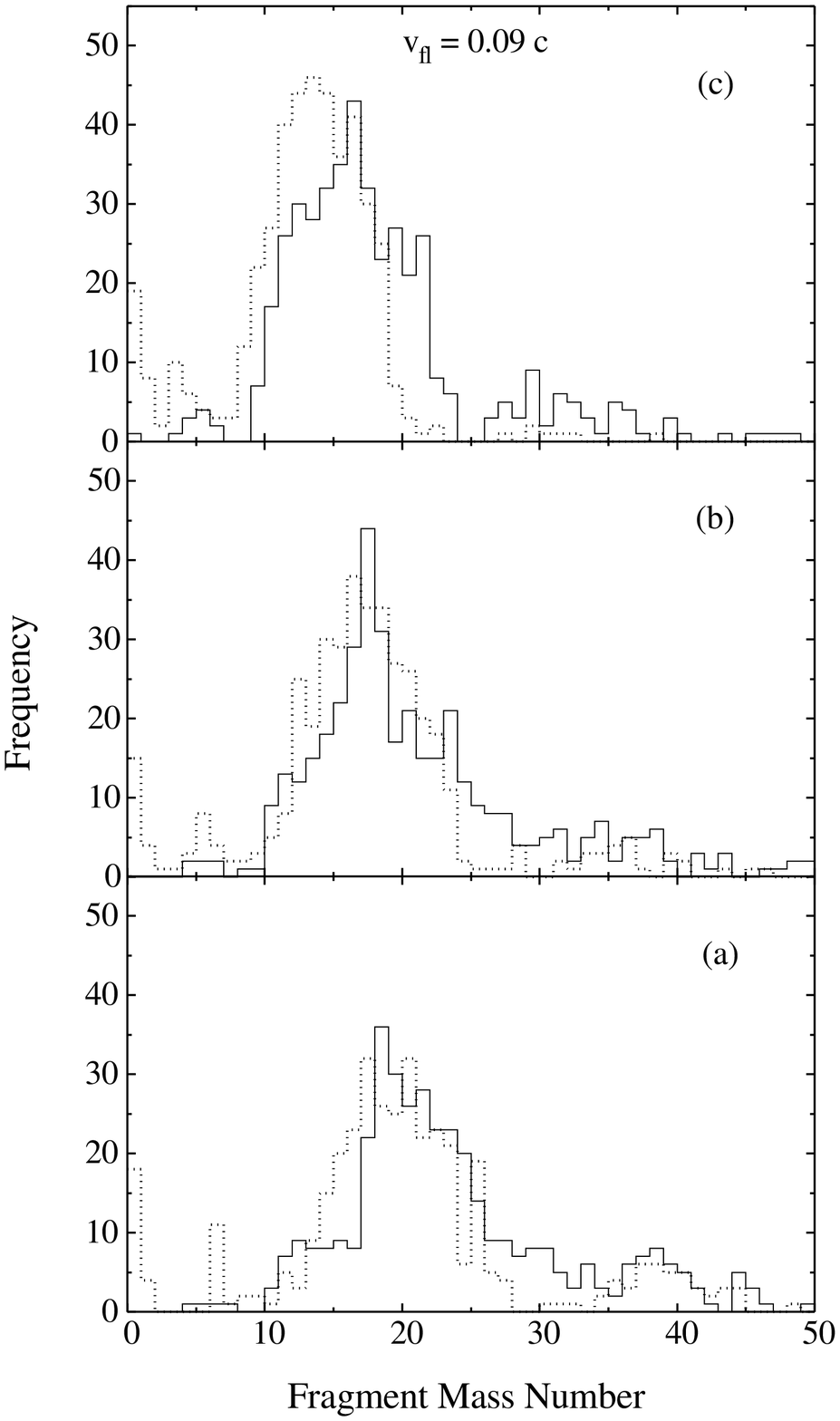,width=0.8\textwidth}
 \vspace{-1.0cm}
\end{center}

{\bf Fig.5} \\
Time evolution of fragment mass abundances for $v_{fl}=$ 0.09c 
shown by dotted histogram in Fig. 4 are compared with 
that calculated from the simulation with the choice of $\Delta p=$ 58 MeV/c and 
$\Delta x={1\over 3}$ fm (shown by solid histogram). 
\end{figure}

\newpage
\begin{figure}
\begin{center}
\epsfig{file=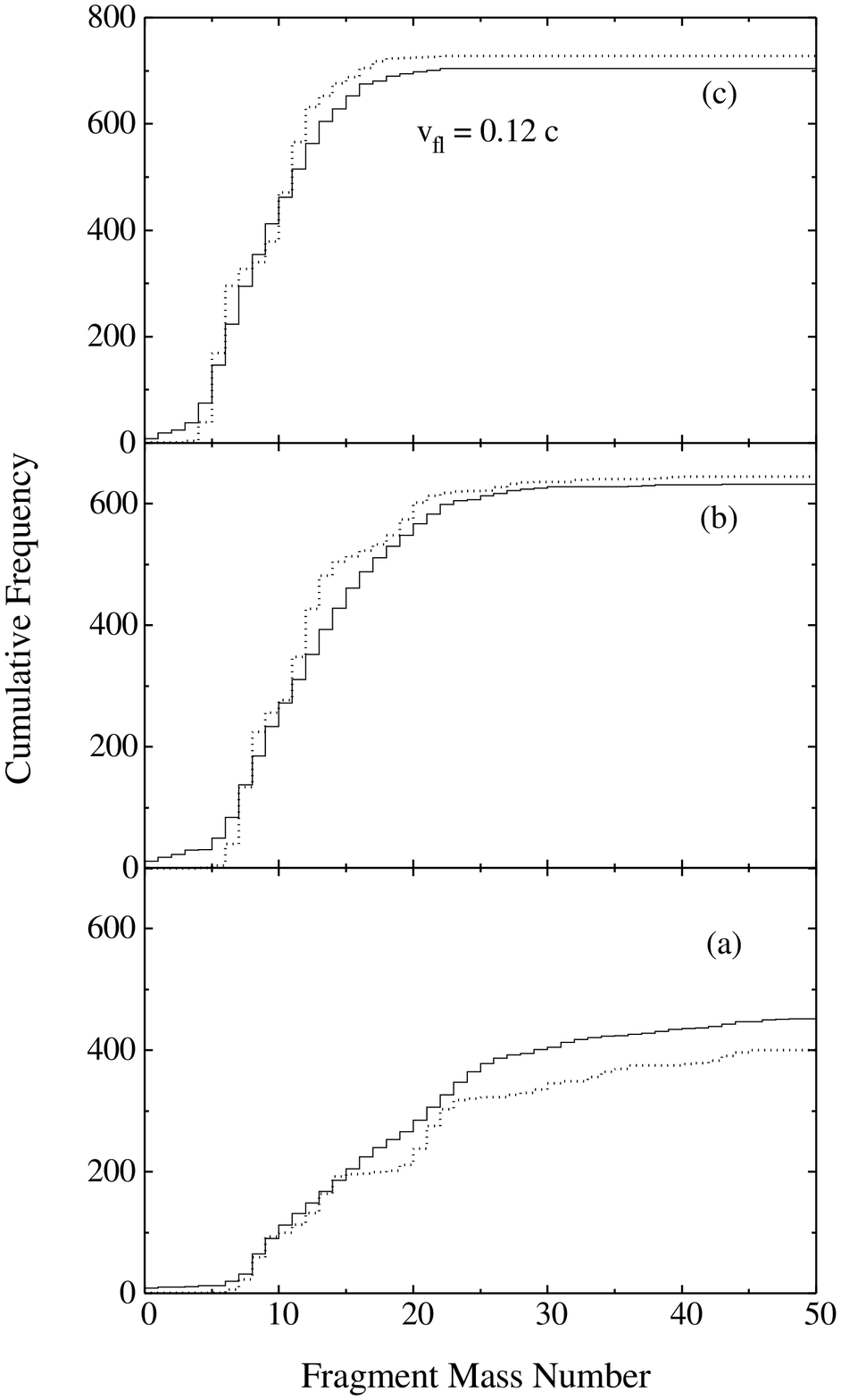,width=0.8\textwidth}
 \vspace{-1.5cm}
\end{center}

{\bf Fig.6} \\
The cumulative frequency versus fragment mass number is plotted 
for different times. Solid histogram represents the same data as shown for 
initial values of $v_{fl}=$ 0.12c in Fig.4 which are compared with the 
calculated results from the simulation 
with the choice of $\Delta p$= 58 MeV/c and $\Delta x={1\over 3}$ fm 
(shown by dotted histogram). 
\end{figure}

\newpage
\begin{figure}
\begin{center}
\epsfig{file=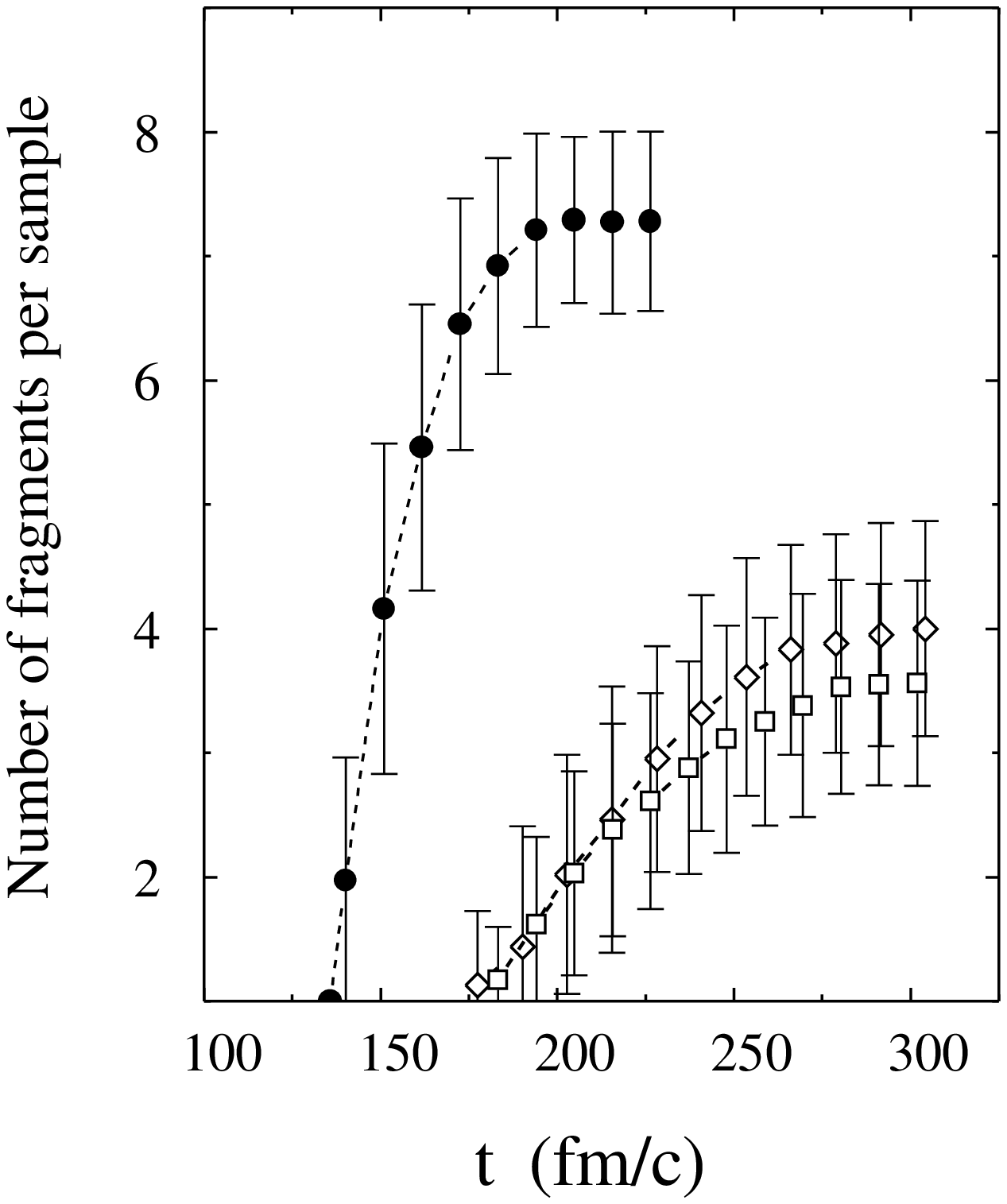,width=0.8\textwidth}
 \vspace{-1.5cm}
\end{center}

{\bf Fig.7} \\
Time evolution of average number of fragments per sample for two 
given values of $v_{fl}=$ 0.12c and 0.09c are shown by solid and open points. 
The open square and open box represent the simulation results for two 
given choices of $\Delta{\bf s}$, ($\Delta p=$ 74 MeV/c, 
$\Delta x={1\over 2}$ fm) and ( $\Delta p=$ 58 MeV/c, $\Delta x={1\over 3}$ fm ) 
respectively. 
The sample fluctuations are indicated by the error bars. The curves connecting 
the points are to guide the eye. 
\end{figure}

\end{document}